\documentclass[twocolumn]{revtex4}

\usepackage{graphicx}  

\usepackage{bm}        
\usepackage{amssymb}   

\usepackage{verbatim}

\begin{document}

\title{Direct observation of non-equilibrium spin population in quasi-one-dimensional nanostructures}

\author{T.-M. Chen}
\author{A. C. Graham}
\author{M. Pepper}
\altaffiliation{Current address: Department of Electronic and Electrical Engineering, University College London, U.K.}
\author{I. Farrer}
\author{D. Anderson}
\author{G. A. C. Jones}
\author{D. A. Ritchie}

\affiliation{
Cavendish Laboratory, J J Thomson Avenue, Cambridge CB3 0HE, United Kingdom
}

\begin{abstract}
Observation of the interplay between interacting energy levels of two spin species is limited by the difficulties in continuously tracking energy levels, and thus leaves spin transport in quantum wires still not well understood. We present a dc conductance feature in the non-equilibrium transport regime, a direct indication that the first one-dimensional subband is filled mostly by one spin species only. How this anomalous spin population changes with magnetic field and source-drain bias is directly measured. We show the source-drain bias changes spin polarisation in semiconductor nanowires, providing a fully electrical method for the creation and manipulation of spin polarization as well as spin-polarized currents.
\end{abstract}

\maketitle
Quantum dynamics of spins has recently been of central interest for a possible technology of spintronics, the main challenge of which is to generate, manipulate, and detect spin-polarised currents in nanostructures \cite{Wolf_science01,Zutic_RMP04}. Realization of this technology principally relies on ferromagnetic contacts \cite{Jedema_Nature01,Urech_NL06}, but efficient spin injection as well as the manipulation of the injected spins remains a challenge for practical applications. Other approaches such as the spin Hall effect \cite{Kato_Science04,Valenzuela_Nature06} and voltage-induced changes in magnetisation \cite{Eerenstein_Nature06,Maruyama_NatureNanotech09} have thus been proposed and have attracted considerable interest. Here we directly demonstrate that quasi-one-dimensional (quasi-1D) mesoscopic devices, nanowires and quantum point contacts, can form the basis of a fully electrical method for the creation and manipulation of spin polarisation as well as spin-polarised currents without high magnetic fields or ferromagnetic injecting contacts.

Semiconductor nanowires and quantum point contacts \cite{Thornton,Berggren86,vanWees88} are among the simplest mesoscopic devices and have significant applications in quantum information processing as charge sensors \cite{Field93,Petta_Science05,Fujisawa_Science06}. Possible spontaneous spin polarisation, as manifest in the so-called 0.7 anomaly \cite{Thomas96}, was suggested to occur in this strongly correlated quasi-1D system and continues to attract fundamental interest within the equilibrium (ohmic) transport regime. Recently, a robust conductance feature at $0.25(2e^2/h)$ in the non-equilibrium transport regime was linked to a pure spin-polarised current \cite{Chen_APL08}, which is a possible candidate for efficient spin injection and detection.

The study of the underlying mechanism of these puzzling conductance features in quasi-1D devices and any associated spin polarisation $P_{1D}=(n_\downarrow - n_\uparrow)/(n_\downarrow + n_\uparrow)$ is severely limited by the difficulties in tracking the quasi-1D subbands, where $n_\downarrow $ and  $n_\uparrow$ are the total densities in each of the two spin bands. Electron transport is frequently studied by measuring the differential conductance $G_{ac}=dI/dV_{sd}$ when a small ac signal of a few microvolts is applied. However, the differential conductance only provides information on the transport properties at a specific value of the chemical potential, detecting the 1D energy level when it moves across a chemical potential and $G_{ac}$ exhibits a change in value, not elsewhere, as $G_{ac}$ remains unchanged, forming a quantised plateau. Measurement of $G_{ac}$ is also performed in the non-equilibrium transport regime, where a relatively large source-drain dc bias comparable to the subband energy spacings (i.e., a few millivolts) is applied to split the chemical potential into two separate source ($\mu_s$) and drain potentials ($\mu_d$), allowing observation of a subband crossing the source and drain chemical potentials only.

A recently developed method of analysis, the dc conductance of the nanowire $G_{dc}=I/V_{sd}$, provides a means to track the subband energy level as it is continuously filled by electrons, consequently giving new insights into the conductance anomalies and electron-electron interactions \cite{Chen_APL08,Chen_PRB09}. The dc conductance --- equivalent to an integration of ac conductance --- of a single spinless subband is given by $G_{dc}=(e^2/h)[\Delta E/eV_{sd}]$, where $\Delta E$ is the energy separation of $\mu_s$ and the bottom of the subband when the subband energy lies between two potentials \cite{Chen_APL08}. Hence, the subband filling $\Delta E$ can be directly tracked during the experiment by measuring $G_{dc}$; it is being able to follow subband movements which gives the technique its usefulness.

In other words, the dc conductance is the integral of information given by the ac conductances over the range of $V_{sd}$, which measures the subband filling energy. Such energy-related information cannot be seen in the traces of $G_{ac}$. In the non-equilibrium regime $G_{dc}$ indicates filling energy and behaves very differently from its corresponding $G_{ac}$, whereas in the equilibrium regime (as $V_{sd} \rightarrow 0$) there is little difference between them.

In this letter, we describe dc conductance measurements in the non-equilibrium transport region, and present a dc conductance anomaly belonging to the 0.7 family. This $G_{dc}$ anomaly directly demonstrates an unusual spin population behaviour of the quasi-1D subbands, viz., the minority spin-up subband almost stops being filled as carrier density increases with split-gate voltage $V_g$. This behaviour was previously suggested \cite{Lassl,Abi_PRB07}, but has not yet been directly observed. How it changes with source-drain bias in the non-equilibrium regime is even much less understood. We now demonstrate how this anomalous subband filling changes as a function of magnetic field and source-drain bias, and provide insight into the puzzling conductance anomaly at $G_{ac} \sim 0.85(2e^2/h)$ \cite{Thomas_Phil98,Kristensen_PRB00}. In addition, we show that the source-drain bias changes the spin polarisation of electrons moving through the quasi-1D devices. The spin-polarised state in the non-equilibrium region is more robust than that in the equilibrium region, and is thus more practical for spintronics. Although the underlying physics is still not understood, we now have a technical ability to electrically generate, manipulate, and detect spin polarisation as well as spin-polarised currents.

\begin{figure}[tbp]
\begin{center}
\includegraphics[width=1\columnwidth]{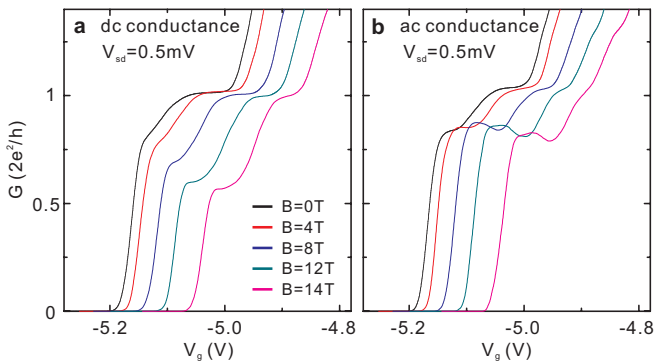}
\end{center}
\caption[Magnetic field dependence of dc and ac conductance at $V_{sd}=0.5$~mV.]
{\small \textbf{Magnetic field dependence of dc and ac conductance at \textit{V$_{sd}$}=0.5$~$mV.} \textbf{(a)} $G_{dc}$ as a function of $V_g$ at $V_{sd}=0.5$~mV and $T=130$~mK for magnetic field $B=0 \rightarrow 14$~T. Traces for different magnetic fields are offset for clarity. \textbf{(b)} $G_{ac}$ measured simultaneously with $G_{dc}$ shown in (a).} 
\label{pin402}
\end{figure}

Figure~\ref{pin402}(a) and (b) respectively show $G_{dc}$ and $G_{ac}$ versus $V_g$ at $V_{sd}=0.5$~mV for in-plane magnetic fields $B=0 \rightarrow 14$T. A shoulder-like feature of $G_{dc}\sim0.8(2e^2/h)$ is found at $B=0$, evolving into a well-defined plateau with increasing magnetic field---the dc conductance value of which drops to $G_{dc}=0.57(2e^2/h)$ at $B=14$~T. In contrast, the corresponding $G_{ac}$, as shown in Figure~\ref{pin402}(b), exhibits a plateau around $0.85(2e^2/h)$ that barely changes with increasing magnetic field. 

The magnetic field induced evolution of the shoulder-like feature in $G_{dc}$ to an almost fully spin polarised state implies that this dc conductance anomaly belongs to the 0.7 family. It is important to note that the source-drain bias alters the spin polarisation. Although both the shoulder-like features in $G_{dc}$ and the 0.7 structure in the equilibrium regime are observed to evolve into a fully spin-polarised $0.5(2e^2/h)$ plateau with increasing magnetic field, the evolution of each of them is slightly different. The dc conductance value of the $G_{dc}$ plateau at $B=14$~T is still larger than the $0.5(2e^2/h)$, showing it is still partially spin-polarised (i.e., $P_{1D}<1$), whereas the 0.7 structure at $V_{sd} \sim 0$ has already evolved into a fully spin-polarised plateau ($P_{1D}=1$) at $B=8$~T (not shown).

\begin{figure}[tbp]
\begin{center}
\includegraphics[width=0.95\columnwidth]{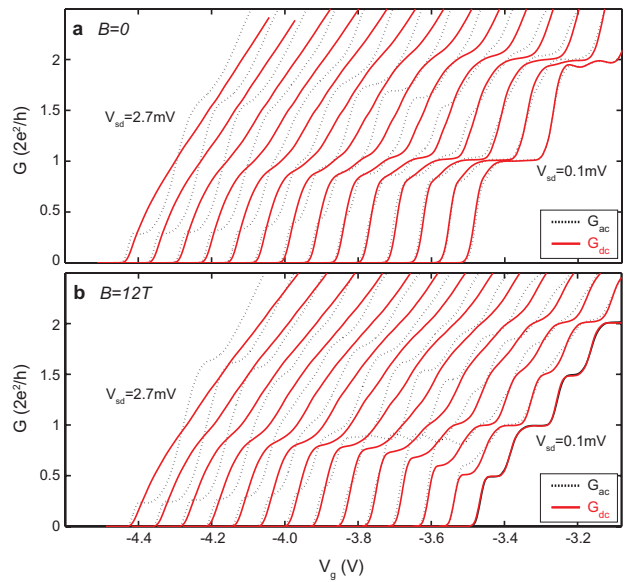}
\end{center}
\caption[Bias dependence of dc and ac conductance.]
{\small \textbf{Bias dependence of dc and ac conductance.} \textbf{(a)} $G_{ac}$ (black dotted) and $G_{dc}$ (red solid) versus gate voltage for a source-drain dc bias from $0.1$~mV to $2.7$~mV at $B=0$ and $T=130$~mK. For clarity, successive traces have been horizontally offset. \textbf{(b)} Same as (a), but at $B=12$~T.} 
\label{pin403}
\end{figure}

Figure~\ref{pin403}(a) shows this $G_{dc}$ feature at incremental values of source-drain bias $V_{sd}$. $G_{dc}$ and $G_{ac}$ are both quantised in multiples of $2e^2/h$ at low $V_{sd}$. At a source-drain bias greater than the energy of the subband spacing, there is always at least one subband energy lying between $\mu_s$ and $\mu_d$, resulting in the absence of quantised plateaux in $G_{dc}$, whereas quantised features in $G_{ac}$ still occur. In addition, it was found that a shoulder-like feature in $G_{dc}$ forms just when $G_{ac}$ evolves into the $0.85(2e^2/h)$ plateau. Figure~\ref{pin403}(b) furthermore shows that these $G_{dc}$ anomalies become more pronounced at $B=12$~T, particularly a non-quantised $G_{dc}$ plateau rising from $0.5(2e^2/h)$ to $0.8(2e^2/h)$, going from low to high dc bias. The $G_{ac}=0.25(2e^2/h)$ anomaly is present for both $B=0$ and $B=12$~T, indicating a zero magnetic field spin polarisation.

\begin{figure}[tbp]
\begin{center}
\includegraphics[width=80mm]{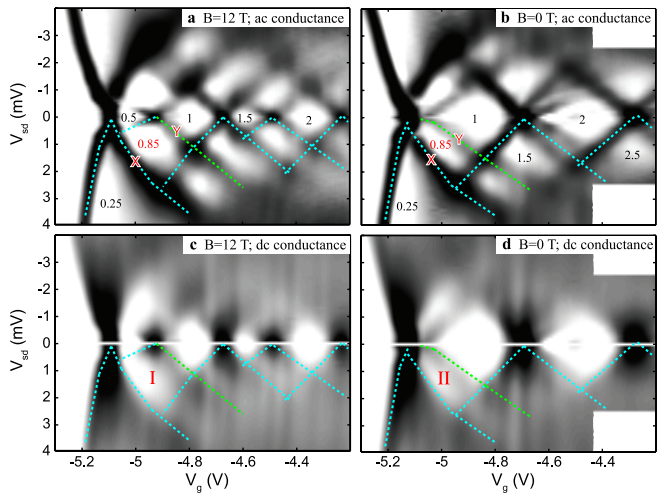}
\end{center}
\caption[Source-drain spectroscopy for ac and dc conductance.]
{\small \textbf{Source-drain spectroscopy for ac and dc conductance.} \textbf{(a)} Grayscale plot of $dG_{ac}/dV_g$ at $B=12$~T versus $V_g$ and $V_{sd}$. Plateaux are white regions, separated by dark branches representing a higher value in $dG_{ac}/dV_g$ as subbands pass through $\mu_s$ or $\mu_d$. Dotted lines are drawn schematically to follow these branches. Differential conductance value of each plateau is marked in units of $2e^2/h$. \textbf{(b)} Grayscale plot of $dG_{ac}/dV_g$ at $B=0$; branches \textbf{X} and \textbf{Y} have evolved from those in (a) at $B = 12$~T. The extra branch, labelled with \textbf{Y}, separating the $2e^2/h$ plateau and the $0.85(2e^2/h)$ plateaux corresponds to a spin-up subband moving across $\mu_d$, implying a ferromagnetic phase at $B = 0$. \textbf{(c),(d)} Grayscale plots of $dG_{dc}/dV_g$, wherein $G_{dc}$ is measured simultaneously with $G_{ac}$ in (a) and (b), respectively. An unexpected dc conductance plateau occurs in the regions, marked with \textbf{I} and \textbf{II}, where a subband still lies between $\mu_s$ and $\mu_d$.} 
\label{Fig3}
\end{figure}

The derivatives of the dc and ac conductance shown in grayscale plots as a function of $V_{sd}$ and $V_g$ [Figure~\ref{Fig3}] give a clearer picture of the behaviour of $G_{dc}$ and the corresponding $G_{ac}$. Black indicates areas of high transconductance, such that the branches in the grayscale plot of $dG_{ac}/dV_g$ [Figure~\ref{Fig3}(a) and (b)] represent a subband edge just passing through a chemical potential, whereas plateaux are white regions. Dotted lines following these branches are drawn schematically in both $dG_{ac}/dV_g$ and $dG_{dc}/dV_g$ grayscale plots in order to illustrate the configuration of the subband levels and the source and drain potentials in various regions. 

At $B=12$~T when the spin degeneracy is clearly lifted, the $G_{ac} \sim 0.85(2e^2/h)$ plateau lies in the region in which the spin-down subband has passed through $\mu_d$ at line \textbf{X}, whilst the spin-up subband lies between $\mu_s$ and $\mu_d$, and passes through $\mu_d$ at line \textbf{Y}, which separates the $0.85(2e^2/h)$ region from the $1\times(2e^2/h)$ plateau, shown in Figure~\ref{Fig3}(a). As the magnetic field is decreased, line \textbf{X} and line \textbf{Y} gradually move closer to each other, but they do not merge with each other even at $B=0$ [Figure~\ref{Fig3}(b)]; meanwhile, the $0.85(2e^2/h)$ plateau itself barely changes with magnetic field. We therefore conclude that the first 1D subband is spin-split even at $B=0$ in the $0.85(2e^2/h)$ plateau region, where the spin-up subband lies \textit{between} $\mu_s$ and $\mu_d$, together with the spin-down subband below $\mu_d$.

Because measurements of $G_{dc}$ as a function of $V_g$ indicate how electrons between $\mu_s$ and $\mu_d$ continuously populate the subband as the 1D channel widens with split-gate voltage, $G_{dc}$ forms a plateau only after the subband moves below both chemical potentials, very different from $G_{ac}$ which forms a plateau whenever the subband energy level is not near a chemical potential. The grayscale plots of $dG_{dc}/dV_g$ [Figure~\ref{Fig3}(c) and (d)] clearly show that, as expected, $G_{dc}$ plateaux occur when there are no 1D subbands lying between two chemical potentials, such as the regions in which the corresponding $G_{ac}$ exhibits the $1\times(2e^2/h)$ and the $2\times(2e^2/h)$ plateau. No plateau is expected in $G_{dc}$ whenever a subband lies between $\mu_s$ and $\mu_d$ (i.e., the region between a pair of V-shaped $dG_{ac}/dV_g$ branches) even though the corresponding $G_{ac}$ does exhibit a plateau, for example, the regions of the $0.25(2e^2/h)$, the $1.5(2e^2/h)$, and the $2.5(2e^2/h)$ plateau in Figure~\ref{Fig3}(d). There is, however, an exception for the region where the $G_{ac}=0.85(2e^2/h)$ plateau is found: a dc conductance plateau occurs when the first spin-up subband still lies between $\mu_s$ and $\mu_d$, implying an unusual population behaviour of this subband.

The 0.7-like dc conductance anomaly in the non-equilibrium regime is a direct indication that the first spin-up subband almost stops populating as the carrier density increases with split-gate voltage, since the value of the dc conductance is equivalent to the subband filling $\Delta E$. In other words, a plateau or shoulder-like feature in $G_{dc}$ as a function of $V_g$ indicates a slower rate at which electrons fill the subband when it is between $\mu_s$ and $\mu_d$. This implies that the 0.7 structure itself is caused by a similar behaviour of a spin-up subband as $V_{sd} \rightarrow 0$, in agreement with the theoretical prediction \cite{Lassl} and the phenomenological models \cite{Bruus,Reilly}. Note that the densities of the two spin types as well as the spin polarisation can be obtained provided that the filling energies $\Delta E$ in each of the two spin subbands are measured.

\begin{figure}[tbp]
\begin{center}
\includegraphics[width=0.8\columnwidth]{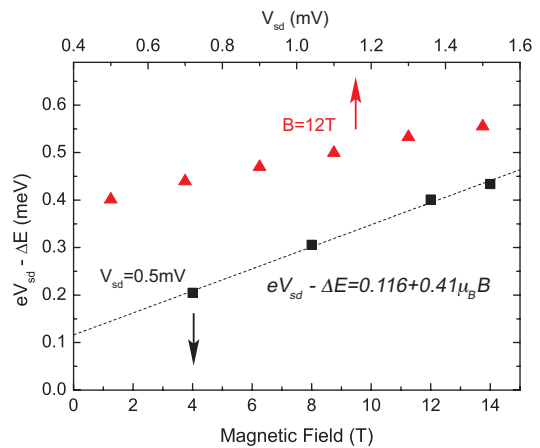}
\end{center}
\caption[Energy difference obtained from the dc conductance.]
{\small \textbf{Energy difference obtained from the dc conductance.} Energy difference between the edge of the first spin-up subband and $\mu_d$, $V_{sd}-\Delta E$, as a function of (i) magnetic field at $V_{sd}=0.5$mV (black symbols at the bottom), obtained from Figure~\ref{pin402}(a), and (ii) source-drain bias $V_{sd}$ at $B=12$T (red symbols at the top), obtained from Figure~\ref{pin403}(b).} 
\label{pin405}
\end{figure}

The conductance value of the $G_{dc}$ anomaly decreases with increasing magnetic field [Figure~\ref{pin402}(a)] on the one hand, and increases with increasing source-drain bias [Figure~\ref{pin403}(b)] on the other. This can be further investigated by calculating $\Delta E$ from the value of $G_{dc}$. Figure~\ref{pin405} shows the energy difference between the edge of the first spin-up subband and $\mu_d$, i.e., $eV_{sd}-\Delta E$, as a function of (i) magnetic field at $V_{sd}=0.5$~mV (black symbols at the bottom), and (ii) source-drain bias $V_{sd}$ at $B=12$~T (red symbols at the top). The points are obtained from the dc conductance value of the anomalous $G_{dc}$ plateau, indicating the energy of the first spin-up level with respect to $\mu_d$ as it stops populating. This energy gap increases linearly with $B$ and follows the relation $eV_{sd}-\Delta E = \alpha +\beta \mu_B B$, where $\alpha$ and $\beta$ are two fitting parameters. The linear-least-square fit gives $\alpha=0.116$~meV and $\beta=0.41$. The parameter $\alpha$ is the energy gap at $B=0$, and $\beta$ is quite close to the effective \textit{g}-factor of bulk GaAs, $|g^*|=0.44$, for the Zeeman splitting. The consistency between them implies that the spin-up subband stops populating further, for a range of gate voltages, just when the first spin-down channel with its energy level near $\mu_d$ is open to electrons moving from the drain. It is important to stress that the nonzero $\alpha$ is equal to the spin gap at $B=0$.

The bias dependence of the energy gap at the top of Figure~\ref{pin405} suggests the same physics. The energy gap, $eV_{sd} - \Delta E$, varying from $0.4$ to $0.55$~meV for source-drain biases between $0.5$~mV and $1.5$~mV at $B=12$~T, is close to the spin-split energy gap, $0.47$~meV, which is obtained using source-drain bias spectroscopy. Note that the energy gap slightly increases with increasing $V_{sd}$, indicating that it is caused by a combination of Coulomb and exchange interactions. There are proposals that this is expected to widen with increasing carrier density \cite{wang96,Berggren02,Lassl,Reilly} and in our case this corresponds to increasing $V_{sd}$ in the $0.85(2e^2/h)$ plateau region.

In the $0.85(2e^2/h)$ plateau region a $G_{ac}$ conductance value of $0.75(2e^2/h)$ is expected rather than the obtained $0.85(2e^2/h)$, although the measured $G_{ac}$ is in general between $0.8(2e^2/h)$ and $0.9(2e^2/h)$. It has been suggested that this anomalous differential conductance value is due to the spin-up subband level being close to $\mu_d$ and consequently filled in part \cite{Kristensen_PRB00}. However, here we have directly measured the subband energy via the dc conductance and found that the subband edge is in fact far above $\mu_d$. The $0.85(2e^2/h)$ plateau behaves in a very different manner from the 0.7 structure, in that it barely changes with temperature up to $1$~K, which indicates that it is not significantly affected by thermal broadening and so the 1D subband edges are not close to the chemical potentials.

A nonlinear electron filling of the spin-up subband as a function of $V_{sd}$ could be responsible for the enhancement of differential conductance of the ``0.85 plateau''. In a non-interacting model for establishing unidirectional dynamics \cite{Glazman,martinmoreno}, the 1D energy level moves linearly with increasing $V_{sd}$, resulting in a quantised $G_{ac}$ of $0.25(2e^2/h)$ for a spinless 1D mode. In contrast, any extra population of electrons, caused by the 1D energy level moving downward instead of being fixed with respect to $(\mu_s + \mu_d)/2$ as $V_{sd}$ is increased, will enhance the differential conductance. This model is consistent with the spin population behaviour observed via the dc conductance anomaly. We note that the 0.85 plateau is observed at $G_{ac}=0.75(2e^2/h)$ in InGaAs \cite{PSimmonds}, implying the significance of the interaction effect in these two systems is different.

To summarize, an anomalous dc conductance feature has been directly related to a non-equilibrium spin population behaviour. This also gives rise to the corresponding differential conductance finite-bias feature at around $0.85(2e^2/h)$. We have investigated this dc conductance anomaly and have shown how the population of the minority up-spins and the spontaneous spin polarisation changes as a function of magnetic field and source-drain bias. This provides a key to a more complete understanding of Coulomb and exchange interactions and the 0.7 anomaly in quasi-one-dimensional systems, as well as a fully electrical method for creation and manipulation of spin-polarised currents.


\section*{Acknowledgements}
We thank useful discussion with F. Sfigakis and L. W. Smith. This work was supported by EPSRC, U.K.

\section*{Methods}
This work utilized GaAs/Al$_{0.33}$Ga$_{0.67}$As heterostructure in which the two-dimensional electron gas is 96~nm below the surface, with a low temperature mobility of $3.97 \times 10^6$~cm$^2$/Vs and a carrier density of $3.37 \times 10^{11}$~cm$^{-2}$. Split-gate devices with a width of 0.8~$\mu$m and lengths from 0.3~$\mu$m to 1~$\mu$m were fabricated and measured and all exhibited similar characteristics. Measurements of differential conductance $G_{ac}=dI/dV_{sd}$ via the variation of $V_{sd}$ due to a small ac excitation voltage of $5~\mu$V, and dc conductance $G_{dc}=I/V_{sd}$ via a larger source-drain dc bias ($V_{sd}$) were performed simultaneously in a dilution refrigerator.

\end{document}